\documentclass[nofootinbib, twocolumn,
 amsmath,amssymb,
 aps,
]{revtex4-2}

\usepackage{graphicx}
\usepackage{dcolumn}
\usepackage{xcolor}
\usepackage{bm}
\usepackage{hyperref}

\newcommand{\Mpl}{M_{\rm Pl}}

\newcommand{\gev}{{\,\rm GeV}}

\usepackage{booktabs} 
\usepackage{tikz}
\usetikzlibrary{arrows.meta,positioning,calc}
\tikzset{
  pq/.style={circle, draw, thick, minimum size=7mm, inner sep=0pt},
  gauge/.style={rectangle, draw, thick, rounded corners=2pt, minimum width=8mm, minimum height=5mm, inner sep=1pt},
  link/.style={thick},
  qlink/.style={thick, -{Latex[length=2mm]}},
  lab/.style={font=\small},
}

\begin{document}

\preprint{APS/123-QED}

\title{Big Axions}

\author{Hannah Banks}
 \email{hannah.banks@nyu.edu}
\affiliation{%
 Center for Cosmology and Particle Physics, Department of Physics, New York University, New York, NY 10003, USA
}
\author{Marius Kongsore}%
 \email{mkongsore@nyu.edu}
\affiliation{%
 Center for Cosmology and Particle Physics, Department of Physics, New York University, New York, NY 10003, USA
}
\author{Neal Weiner}
 \email{neal.weiner@nyu.edu}
\affiliation{%
 Center for Cosmology and Particle Physics, Department of Physics, New York University, New York, NY 10003, USA
}%

\date{\today}

\begin{abstract}
We introduce \textit{big axions}: axion models in which a Nambu--Goldstone mode emerges from the collective spontaneous breaking of a network of $U(1)$ symmetries delocalized in theory space. Big axions naturally realize high-quality accidental global symmetries, admit both pre- and post-inflationary cosmological histories, and exhibit rich topological structures that interpolate between ordinary Peccei--Quinn axions and axions which descend from extra-dimensional gauge fields. We identify a minimal phenomenologically viable subclass, \textit{little big axions}, and demonstrate that they provide a robust solution to the strong charge--parity problem in quantum chromodynamics while potentially accounting for some or all of the dark matter of the universe.
\end{abstract}

\maketitle


\section{\label{sec:introduction}Introduction}

Light scalar fields have been both a blessing and a curse in modern particle physics. They lie at the heart of solutions to a number of important problems---giving mass to the gauge bosons of the Standard Model (SM)~\cite{Englert:1964et,Higgs:1964pj,Guralnik:1964eu}, addressing the strong charge--parity (CP) problem~\cite{Peccei:1977hh,Peccei:1977ur,Weinberg:1977ma,Wilczek:1977pj}, driving inflation~\cite{Guth:1980zm,Linde:1981mu,Albrecht:1982wi}, and providing a natural candidate for the dark matter (DM) of the universe~\cite{Preskill:1982cy,Abbott:1982af,Dine:1982ah}, among others. Their utility, however, is hampered by our understanding of naturalness, which suggests that scalar fields should generically have a mass near their cutoff~\cite{Susskind:1978ms,tHooft:1979rat}.

An elegant resolution of this tension is for light scalar fields to appear as Nambu--Goldstone bosons of a broken exact, or nearly exact, global symmetry~\cite{Nambu:1960tm,Goldstone:1961eq,Goldstone:1962es}. The shift symmetry that accompanies the scalar field, emerging from the non-linear realization of this global symmetry~\cite{Weinberg:1968de,Salam:1969rq,Coleman:1969sm,Callan:1969sn}, protects its mass while simultaneously forbidding non-derivative interactions. Such particles provide natural solutions, in particular to the strong CP problem in the form of the axion~\cite{Peccei:1977hh,Peccei:1977ur,Weinberg:1977ma,Wilczek:1977pj}.

This simple idea, however, fades when confronted by cosmology and ideas arising from quantum gravity. Spontaneously broken global symmetries often yield topological structures, such as cosmic strings and domain walls, that are excluded by even the most basic cosmological observations~\cite{Zeldovich:1974uw,Sikivie:1982qv,Vilenkin:1984ib}. At the same time, quantum gravity challenges the very premise that a global symmetry can persist with sufficient quality to admit a functional Nambu--Goldstone boson~\cite{Giddings:1988cx,Kamionkowski:1992mf,Holman:1992us,Barr:1992qq,Ghigna:1992iv,Banks:2010zn,Harlow:2018tng}.

One approach to addressing these problems is to protect the global symmetry using a discrete gauge symmetry. This can arise by spontaneously breaking a $U(1)$ gauge symmetry with a very large charge, leaving a remnant $\mathbb{Z}_N$ gauge symmetry which is indistinguishable from a high-quality discrete global symmetry within the low-energy effective theory~\cite{Krauss:1988zc,Lazarides:1985bj,Dias:2002gg,Harigaya:2013vja,Fukuda:2017ylt}. Alternatively, one may view the Nambu--Goldstone boson as a component of a higher-dimensional gauge field, with the shift symmetry embedded in a global higher-form symmetry~\cite{Witten:1984dg,Arkani-Hamed:2003xts,Choi:2003wr,Svrcek:2006yi,Reece:2023czb,Reece:2025thc,Craig:2024dnl}. This embedding renders the global symmetry more robust, and many of the problematic phenomenological implications of generic global symmetries are absent.

Deconstructed models~\cite{Arkani-Hamed:2001kyx,Hill:2000mu,Arkani-Hamed:2001wsh} aim to offer a four-dimensional realization of certain features of higher-dimensional gauge theories. By placing gauge fields at sites in a moose or quiver diagram and using  link fields to Higgs the theory down to subgroups, many properties of higher-dimensional theories can be reproduced~\cite{Georgi:1985hf,Douglas:1996sw}. In particular, non-locality in the extra dimensions can be represented by non-locality in theory space or equivalently by collective symmetry breaking, with dangerous operators emerging only through collective insertions involving many fields and therefore remaining highly suppressed~\cite{Arkani-Hamed:2001nha,Arkani-Hamed:2002ikv}.

In this Letter, we explore the idea of global theory space protection for  light Nambu--Goldstone bosons. Inspired by the lessons of deconstruction, we elevate the structure of theory space from a tool for discretizing extra dimensions to an organizing principle for protected low-energy physics. To this end, we formulate a broad class of axion theory spaces characterized by an integer charge matrix $\hat{\mathbf{Q}}$, whose global properties determine the existence and structure of the protected low-energy degrees of freedom. This framework encompasses a diverse landscape of axion theory space architectures that extends significantly beyond previously considered one-dimensional moose models, to reveal novel algebraic and topological features. For concreteness, we focus on a particularly illuminating subclass that admits a graphical dual moose interpretation in which gauge fields are represented as links and Higgs fields as sites. The resulting protected modes are `big axions': collective Nambu–Goldstone modes composed of many Higgs fields whose shift symmetries can only be violated by high-dimension operators that span the network, thereby ensuring quality. We explore how these Nambu-Goldstones can realize the axion of quantum chromodynamics (QCD) and find that, in the most minimal DFSZ-/KSVZ-like models~\cite{Kim:1979if,Shifman:1979if,Dine:1981rt,Zhitnitsky:1980tq}, the domain wall number~\cite{Sikivie:1982qv} is naturally unity whilst the quality of the axion is tied to the number of fermions that carry the anomalous Peccei-Quinn (PQ) charge. Remarkably, these models are compatible with both pre- and post-inflationary cosmological histories, the latter being a feature shared with few high-quality axion models~\cite{Lu:2023ayc,Petrossian-Byrne:2025mto,Loladze:2025uvf}. In addition, we find that many existing high-quality solutions, such as that of Barr and Seckel~\cite{Barr:1992qq} and one-dimensional gauged U(1) moose models~\cite{Hill:2002kq,hor2026deconstructingextradimensionalaxion}, including variants with clockwork-style charge assignments\footnote{Note that other clockwork axions constructed from chains of \textit{global} U(1) symmetries have also been proposed~\cite{Kaplan:2015fuy,Choi:2015fiu}. The primary goal of these models is to engineer localized zero modes and thereby generate hierarchical effective couplings. Since the underlying clockwork symmetry is global, they do not address the axion quality problem by themselves.} that result in hierarchical zero mode profiles ~\cite{Ahmed:2016viu,Coy:2017yex,Bonnefoy:2018ibr}, emerge as special cases within the broader big axion framework.

We lay out this Letter as follows. We begin in Sec.~\ref{sec:big-axions} by introducing the big axion and its connection to the kernel of the charge matrix $\bf{\hat{Q}}$. In Sec.~\ref{sec:big-qcd-axions}, we realize scenarios in which it acts as the QCD axion, before constructing minimal phenomenologically viable models of big QCD axions in  Sec.~\ref{sec:the-littlest-big-axion}. Finally, in Sec.~\ref{sec:outlook}, we discuss possible future directions, including a deconstruction-based understanding of extra-dimensional axions, mixed gauge-global string cosmology, connections to flavor physics, quintessence, and more.

\section{\label{sec:big-axions}Big Axions}

In this Section, we introduce the big axion framework. We first define the big axion charge matrix and illustrate how it controls both the structure and quality of any big axion. We then show how a big axion may be represented by a graph and highlight important big axion subclasses. Finally, we discuss the criteria under which big axions can be viewed as deconstructions.
\subsection{The Big Axion and its Theory Space}
A big axion theory space is built from $N_{\Phi}$ complex scalar fields $\Phi_i$, $i \in \{1,..,N_{\Phi}\}$, henceforth referred to as \textit{network Higgses}, each of which possesses a spontaneously broken $U(1)$ phase rotation symmetry. The network Higgses carry charge under $N_{A}$ $U(1)_\alpha$ gauge fields\footnote{In principle, these fields could arise from the Cartan subgroup of a non-Abelian group.} $A_{\alpha}$, $\alpha \in \{1,..,N_{A}\}$, which effectively gauge some linear combinations of the  phase rotation symmetries while leaving others global. The gauge charges of the network Higgs fields can be encoded in an $N_{A} \times N_{\Phi}$ matrix
\begin{equation}
\hat{\boldsymbol{Q}}: \mathbb{Z}^{N_\Phi} \rightarrow \mathbb{Z}^{N_{A}} \, ,
\end{equation}
whose $(\alpha, i)$ entry corresponds to the charge of the $i^{\rm th}$ network Higgs field under the $U(1)_{\alpha}$ gauge symmetry. This matrix completely determines the gauge structure of the Higgs sector and serves as the central organizing object of the big axion framework.

Big axions themselves emerge as phases of gauge invariant products of network Higgs fields. They are uniquely specified by primitive vectors $\mathbf{n}\in\mathbb{Z}^{N_\Phi}$ lying in the kernel of the charge matrix
\begin{equation}
    \hat{\boldsymbol{Q}}\mathbf{n} = 0\, .
\end{equation}
Each of these vectors defines a gauge invariant monomial operator 
\begin{equation}\label{eq:big-axion-parent}
\mathcal{O}_\mathbf{n} = \prod_{i=1}^{N_{\Phi}} \Phi_i^{[n_i]} \, ,
\end{equation} 
the phase of which corresponds to an axion degree of freedom. Here, the superscript $[n_i]$ is shorthand for $\Phi_i^{|n_i|}$ if $n_i\geq 0$ and $(\Phi_i^*)^{|n_i|}$ if $n_i<0$. Since each of the network Higgses carries charge under the  $U(1)_{\alpha}$ gauge symmetries, only specific products of fields whose charges contrive to cancel under each of the  $U(1)_{\alpha}$  symmetries will be gauge invariant. As a result, both the big axion and its parent Higgs monomial operator must be built from multiple \textit{different} network Higgses, rendering  them `non-local' in theory space. 

These collective structures are governed by the kernel of the charge matrix $\hat{\mathbf{Q}}$: its dimension counts the number of independent big axions, each primitive null vector $\mathbf{n}$  defines a big axion parent operator, and the components $n_i$ delineate the multiplicities of each scalar (or its complex conjugate if $n_i<0$) in the parent monomial operator. The remaining pseudoscalars, of which there are $N_\Phi - \dim ( \ker \hat{\mathbf{Q}})$, are eaten by linear combinations of gauge fields in the big axion network via the Higgs mechanism, giving masses to vector bosons. Any leftover gauge fields, counted by the dimension of the \textit{left} kernel of $\hat{\mathbf{Q}}$, remain as decoupled massless hidden photons in the infrared spectrum unless they are gapped by some other mechanism.

We may explicitly parameterize the network Higgs fields as
\begin{equation}
\Phi_i = \frac{f_i+\rho_i}{\sqrt{2}}e^{i \frac{a_i}{f_i}} \, ,
\end{equation}
where $\rho_i$ is a heavy real scalar field, $a_i$ a pseudoscalar field, and $f_i/\sqrt{2}$ the vacuum expectation value of the network Higgs. The (dimensionless) gauge invariant big axion $\Theta_\mathbf{n}$ is then given by
\begin{equation}
    \Theta_\mathbf{n} \equiv \sum_i^{N_{\Phi}}  \frac{n_i a_i}{f_i} = \sum_i^{N_{\Phi}} n_i \theta_i \, ,
\end{equation}
where we have defined $\theta_i = a_i/f_i$. The canonically normalized big axion can be written as $a_\mathbf{n} = \Theta_\mathbf{n} f_{\rm eff}$, with an effective decay constant given by
\begin{equation}
\frac{1}{f_{\rm eff}^2} =  \sum_{i}^{N_{\Phi}} \left( \frac{n_i}{f_i} \right)^2\, .
\end{equation}

One of the key features of the big axion is that it possesses a high-quality shift symmetry. Quantum gravity effects, expected to explicitly break the big axion shift symmetry at the Planck scale, do so at lowest order via the gauge invariant operator
\begin{equation}\label{eq:quality}
    \mathcal{L} \supset \frac{c\prod_{i}\Phi_{i}^{[n_i]}}{M_\text{Pl}^{(\sum_i |n_i|)-4}} + \text{h.c.}\, ,
\end{equation}
where $c$ is a complex Wilson coefficient with $\mathcal{O}(1)$ modulus and $M_\mathrm{Pl}$ is the Planck mass. Due to the delocalization of the axion across theory space, the operator dimension
\begin{equation}
D_{\rm quality} = \sum_i |n_i| \, ,
\end{equation}
can naturally be large, ensuring the big axion shift symmetry is of high quality. 

\subsection{The Big Axion as a Graph}
Although the big axion charge matrix is the defining element of the model, it can often be interpreted as the incidence matrix for a charge-weighted graph. This offers an alternative, highly intuitive graphical route to constructing big axion theories. To motivate this perspective, we begin by noting that traditional one-dimensional deconstructed models   \`a la Ref.~\cite{Hill:2002kq} correspond to a specific subclass of big axion. In such (de)constructions, each Higgs field is assigned equal and opposite charges under a pair of `adjacent' gauge groups. In terms of the big axion charge matrix, these models correspond to a restricted subset of square matrices (i.e. with $N_{\Phi} = N_A$) which contain exactly two non-zero entries in every row and column. These models are typically formulated in terms of a moose (or quiver) diagram in which the gauge groups are represented as sites in theory space, with the Higgs fields forming nearest-neighbor links connecting them. Gauge-invariant operators correspond to closed cycles around the graph and consequently involve every Higgs field in the construction, making the notion of \textit{theory space non-locality} manifest. 

The conventional moose is too restrictive to represent the full landscape of possible big axion charge matrices however, breaking down should any Higgs carry charge under more than two gauge groups. To generalize beyond this, whilst retaining the intuition afforded by a simple graphical description, we instead introduce a `dual moose' formulation in which the conventional roles of the Higgs and gauge fields are interchanged. In this picture, each Higgs field now corresponds to a graph vertex  and may carry charge under an arbitrary number of gauge symmetries, encoded in the number of incident gauge links. Such high degree vertices enable the construction of graphs, and by proxy theory spaces, with highly non-trivial topologies.
 
\begin{figure}
    \centering
    \resizebox{\columnwidth}{!}{%
\begin{tikzpicture}[
    site/.style={
        circle,
        draw,
        fill=white,
        minimum size=5.4mm,
        inner sep=0pt,
        font=\scriptsize
    },
    link/.style={line width=0.95pt},
    hiddenlink/.style={line width=0.8pt,dashed,gray!70},
    panellab/.style={font=\bfseries},
    glabel/.style={
        font=\scriptsize,
        inner sep=0.8pt
    },
]

\def\sep{4.5}

\def\panelxmin{-2.55}
\def\panelxmax{ 2.55}
\def\panelymin{-1.85}
\def\panelymax{ 1.85}

\begin{scope}[shift={(0,0)}]
    \path[use as bounding box]
        (\panelxmin,\panelymin) rectangle (\panelxmax,\panelymax);

    \node[panellab] at (-2.28,1.50) {(a)};

    \coordinate (T1) at ( 0.00,  1.18);
    \coordinate (T2) at (-1.35, -0.70);
    \coordinate (T3) at ( 1.35, -0.70);
    \coordinate (T4) at ( 0.28,  0.10);

    \draw[link] (T1) -- (T4);

    \draw[link] (T1) -- (T2);
    \draw[link] (T2) -- (T3);
    \draw[link] (T3) -- (T1);

    \draw[link] (T2) -- (T4);
    \draw[link] (T3) -- (T4);

    \node[glabel] at (-0.92, 0.35) {$A_{1}$};
    \node[glabel] at ( 0.90, 0.35) {$A_{2}$};
    \node[glabel] at ( 0.00,-0.89) {$A_{4}$};

    \node[glabel] at ( -0.05, 0.57) {$A_{3}$};
    \node[glabel] at (-0.55,-0.10) {$A_{5}$};
    \node[glabel] at ( 0.68,-0.38) {$A_{6}$};

    \node[site] at (T1) {$\Phi_1$};
    \node[site] at (T2) {$\Phi_2$};
    \node[site] at (T3) {$\Phi_3$};
    \node[site] at (T4) {$\Phi_4$};
\end{scope}

\begin{scope}[shift={(\sep,0)}]
    \path[use as bounding box]
        (\panelxmin,\panelymin) rectangle (\panelxmax,\panelymax);

    \node[panellab] at (-2.28,1.50) {(b)};

    \coordinate (S1) at ( 0.00,  0.00);   
    \coordinate (S2) at ( 0.00,  1.10);
    \coordinate (S3) at (-1.25, -0.90);
    \coordinate (S4) at ( 1.25, -0.90);

    \draw[link] (S1) -- (S2);
    \draw[link] (S1) -- (S3);
    \draw[link] (S1) -- (S4);

    \node[glabel] at ( 0.24, 0.58) {$A_{1}$};
    \node[glabel] at (-0.78,-0.27) {$A_{2}$};
    \node[glabel] at ( 0.78,-0.27) {$A_{3}$};

    \node[site] at (S1) {$\Phi_1$};
    \node[site] at (S2) {$\Phi_2$};
    \node[site] at (S3) {$\Phi_3$};
    \node[site] at (S4) {$\Phi_4$};
\end{scope}

\end{tikzpicture}
}
\vspace{-10mm}
    \caption{Examples of big axion theory spaces. (a) A geometric big axion theory space with charge matrix~\eqref{eq:tetrahedron}, corresponding to a \textit{tetrahedron}. (b) A related accidental big axion theory space with charge matrix~\eqref{eq:star}, corresponding to a reduced tetrahedron or a \textit{star}. With appropriate charge assignments and fermion couplings, discussed in Sec.~\ref{sec:the-littlest-big-axion}, both of these theory spaces provide a high-quality QCD axion candidate. The geometric big axion has additional light gauge bosons relative to the accidental big axion. Additional examples of big axion theory spaces are shown in Appendix~\ref{app:general}.}
    \label{fig:diagrams}
\end{figure}
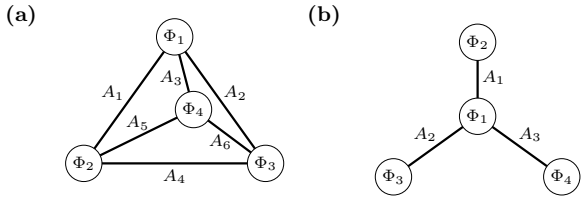

To illustrate this, consider the following charge matrix
 \begin{equation}\label{eq:tetrahedron}
    \mathbf{\hat{Q}}_{\rm tetra} = \begin{pmatrix}
        -1 & q & 0 & 0\\
        -1 & 0 & 1 & 0\\
        -1 & 0 & 0 & 1 \\
        0 & -q & 1 & 0 \\
        0 & -q & 0 & 1 \\
        0 & 0 & -1 & 1
    \end{pmatrix} .
\end{equation}
This corresponds to a tetrahedral dual moose, as depicted in Fig.~\ref{fig:diagrams}(a), and gives rise to a single big axion characterized by the vector $\mathbf{n}^T=(q,1,q,q)$.  Another (related) charge matrix is
\begin{equation}\label{eq:star}
    \mathbf{\hat{Q}}_\mathrm{star} = \begin{pmatrix}
        -1 & q & 0 & 0\\
        -1 & 0 & 1 & 0\\
        -1 & 0 & 0 & 1 \\
    \end{pmatrix} ,
\end{equation}
which maps to a star shaped graph, shown in Fig.~\ref{fig:diagrams}(b). In spite of its smaller theory space, this model yields the same single big axion, $\mathbf{n}^T=(q,1,q,q)$, as the tetrahedron. This is no coincidence. When $N_A\geq N_\Phi$, any big axion is necessarily \textit{geometric} in the sense that the theory space needs to be carefully engineered to prevent all of the Nambu--Goldstone bosons from being eaten by gauge fields. In contrast, when $N_A < N_\Phi$ one obtains  \textit{accidental} big axions whose existence follows automatically from the mismatch between the number of Higgs and gauge degrees of freedom in the network, independent of the theory space geometry. Since any geometric big axion theory space with $N_A\geq N_\Phi$ contains decoupled vector bosons, it is possible to remove the redundant gauge fields by projecting out the left kernel of the charge matrix without impacting the right kernel and hence big axion sector.  This is precisely the relation between Eq.~\eqref{eq:tetrahedron} and Eq.~\eqref{eq:star}, the latter is a reduction of the former. We find that the big axion in both theories remains intact upon extending the network with additional charged Higgs fields, although such extensions may also enlarge the kernel of the charge matrix, giving rise to additional big axionic degrees of freedom. More generally one may identify equivalence classes of big axion theories which, modulo field redefinitions, describe the same physics. The theories in each class  possess the same $N_A$ and $N_{\Phi}$ and have charge matrices which are related by transformations under the product group  $GL(N_A,\mathbb{Z})\times S_{N_\Phi}\times (\mathbb{Z}_2)^{N_\Phi}$, corresponding to, from left to right, linear gauge field redefinitions, permutations of the network Higgs fields, and complex conjugation of the network Higgs fields, respectively.

While the dual moose formulation offers a straightforward and intuitive way to construct big axion theory spaces with non-trivial topologies, it is still not fully general since, by construction, each gauge field can only connect to two sites.  In full generality, $\bf{\hat{Q}}$ can be interpreted as the incidence matrix of a charge-weighted \textit{hypergraph}. Nevertheless the dual moose already accommodates a remarkably rich variety of non-local theory space topologies and suffices to illustrate the essential phenomenology of generic big axion models whilst remaining both intuitive and tractable. For concreteness we shall focus on the charge matrices~\eqref{eq:tetrahedron} and~\eqref{eq:star} throughout the remainder of this work, offering examples of more general constructions in Appendix~\ref{app:general}.

\subsection{The Big Axion as a Deconstruction}
Beyond the features emphasized above, big axions also make contact with several structural aspects of quantum field theory. One particularly intriguing point is that, in special cases, big axions may be regarded as deconstructions of not only flat extra dimensions, but more general manifolds. For example, when $q=1$ in $\mathbf{\hat{Q}}_{\rm tetra}$~\eqref{eq:tetrahedron}, the tetrahedral big axion is effectively a deconstruction of a two-sphere: the charge matrix is the face-to-edge incidence matrix for the boundary of a tetrahedral 3-cell, specifying the orientations of each of the tetrahedron's triangular faces. In this language, the Higgs fields  are better thought of as residing on the \textit{faces} of the tetrahedral complex (or, more accurately, its \textit{dual} complex), so that their phases are the natural lattice variable for a compact Kalb-Ramond two-form gauge field $B_2$, which emerges when the triangular faces can be glued together in such a way that the resultant two-surface has no boundary. The continuum zero modes of such higher-form fields are usually counted by cohomology groups, while in the discretized theory, the same structure appears through the cellular chain complex. For the tetrahedral lattice deconstruction of a sphere, the incidence matrix gives $\mathrm{ker}\,\mathbf{\hat{Q}}_\mathrm{tetra}\simeq H_2(S^2,\mathbb{Z})\simeq \mathbb{Z}$, indicating the presence of a single wrapped two-form axion in the theory. In the continuum limit, the big axion becomes a higher-form holonomy $\Theta_{\bf n} \sim \int_{S^2} B_2$, and what at finite lattice spacing looks like an ordinary zero-form shift symmetry of a collection of scalar phases instead resembles a higher-form symmetry. In this sense, deconstructed big axions \textit{sit between} ordinary PQ axions and axions obtained from Abelian gauge fields wrapping cycles in extra dimensions. More general big axions need not have such a direct manifold interpretation---the charge matrix $\hat{\mathbf{Q}}$ allows for much more general structures---but the deconstructed subclass already exhibits a rich interplay between topology, higher-form gauge fields, and generalized symmetries. A systematic exploration of these connections lies beyond the scope of this Letter; here we instead focus on constructing minimal phenomenologically viable models of big QCD axions.

\section{\label{sec:big-qcd-axions}Big QCD Axions}
Having established the general framework of big axions, we must still address whether such particles can serve as a viable setup to address the strong CP problem. 
In this Section, we shall pursue this, and couple the general class of big axions introduced above to QCD, promoting them to big \textit{QCD axions}. This can be accomplished in a gauge anomaly-free way via two distinct mechanisms:
\begin{enumerate}
    \item Adding anomalous fermion content throughout the Higgs field network such that the multiplicity and couplings of fermions at each site mirrors the structure of the big axion monomial.
    \item Adding anomalous fermion content at a subset of network Higgs field sites, paired with anomaly-canceling Green-Schwarz~\cite{Green:1984sg} terms.
\end{enumerate}
The latter case effectively amounts to introducing a second pseudoscalar that couples to QCD and mixes with a subset of sites in the big axion network. Addressing the quality problem of this new QCD axion in a UV-complete manner is then a separate issue from ensuring the quality of the big axion network: for example, if the Green-Schwarz terms arise from anomaly inflow, extra dimension(s) will ensure quality rather than theory space non-locality. For this reason, we focus on the former mechanism in this work.

We have two choices for which vectorlike quarks to couple to our network Higgs field sites. The first is adding new dynamical beyond-the-SM (BSM) fermions $\Psi$ and $\Psi^c$ in the $\mathbf{3}$ and $\bar{\mathbf{3}}$ of $SU(3)_c$, respectively, to our model and coupling them to the Higgs fields via
\begin{equation}\label{eq:ksvz-coupling}
    \mathcal{L}\supset \lambda_A\Phi_i^*\Psi_A\Psi_A^c + \mathrm{h.c.} \, ,
\end{equation}
where the index $i$ labels the network Higgs site at which the coupling term lives, the index $A$ labels the fermion coupled at that site, and $\lambda_A$ is a real coupling constant.\footnote{Here, we take the complex conjugate of $\Phi_i$ in the coupling in order that the signs of the KSVZ and DFSZ $U(1)_\alpha\times SU(3)_c\times SU(3)_c$ gauge anomaly contributions match.} This is effectively a KSVZ-type mechanism~\cite{Shifman:1979if,Kim:1979if}. Taking only the $\mathbf{3}$ to be charged under the hidden $U(1)_\alpha$ gauge groups, these are the only allowed fermion mass terms. The second choice is to generate the anomalous axion-gluon coupling by coupling the big axion network to the SM quarks in a DFSZ-type mechanism~\cite{Dine:1981rt,Zhitnitsky:1980tq}. This can be done by introducing two electroweak Higgs doublets $H_u$ and $H_d$ with the following couplings to the SM quarks and network Higgs fields
\begin{multline}\label{eq:dfsz-coupling}
    \mathcal{L}\supset \lambda_{ij}\Phi_i \Phi_j H_u H_d + y_u Q^a H_u (u^c)^a\\
    + y_d Q^a H_d (d^c)^a + y_e L^a H_d (e^c)^a + \mathrm{h.c.}\, ,
\end{multline}
where the subscripts $i$ and $j$ label the network sites on which the couplings live, the superscript $a$ indicates the flavor generation, and $\lambda_{ij}$, $y_u$, $y_d$, and $y_e$ are coupling constants.

These choices are not mutually exclusive. To keep QCD asymptotically free while simultaneously maintaining high quality in the big axion network, it is advantageous to employ a hybrid of these two  mechanisms. To one-loop order, the QCD beta function is~\cite{Huston:2023ofk}
\begin{equation}\label{eq:qcd-running}
    \beta(\alpha_s) = \mu^2\frac{d\alpha_s}{d\mu^2} = -\frac{1}{12\pi}\left(33-2 n_f\right)\alpha_s^2 \, ,
\end{equation}
where $\mu$ is the renormalization scale and $n_f$ is the number of quark flavors, i.e. $\mathbf{3}\oplus\bar{\mathbf{3}}$ pairs. Since the SM has six such flavors, the big QCD axion network may contain at most ten additional KSVZ-like Dirac triplet quarks without sacrificing the asymptotic freedom of QCD.\footnote{If these KSVZ-like fermions are sufficiently heavy, even if QCD develops a Landau pole, that pole can exist above $M_\mathrm{Pl}$, rendering the SM coupled to the big axion a perfectly valid effective field theory up to the Planck scale. Nevertheless, we impose asymptotic freedom as a conservative criterion on the theory.}

Since any individual fermion coupled to a site $j$ generates $U(1)_{\alpha}\times SU(3)_c\times SU(3)_c$ gauge anomalies, they must be added throughout the network in such a way that these anomalies cancel. Taking $\mathbf{k}\in \mathbb{Z}^{N_\Phi}$ and $\mathbf{d}\in \mathbb{Z}^{N_\Phi}$ to be vectors indicating the multiplicity of KSVZ and DFSZ quarks coupled to the various sites with couplings given by Eq.~\eqref{eq:ksvz-coupling} and Eq.~\eqref{eq:dfsz-coupling}, respectively, the cancellation of these anomalies is captured by the condition
\begin{equation}\label{eq:qcd-anomaly-cancelation-counting}
    \mathbf{k} + \mathbf{d} = N_\mathrm{DW}\mathbf{n} \, ,
\end{equation}
for some $N_\mathrm{DW}\in\mathbb{Z}$, the big axion domain wall number. A detailed derivation of Eq.~\eqref{eq:qcd-anomaly-cancelation-counting} is carried out in appendix~\ref{app:qcd-anomaly-counting}. An important corollary of this relation is that the total number of quarks in the theory must satisfy
\begin{equation}
    \sum_i \left(|d_i|+|k_i|\right) \geq D_\mathrm{quality} \, ,
\end{equation}
with equality if $N_\mathrm{DW}=1$.\footnote{This assumes there are no additional quarks coupled to the conjugate of the Higgses appearing in Eq.~\eqref{eq:ksvz-coupling} and Eq.~\eqref{eq:dfsz-coupling}, which could neutralize their anomaly contribution and allow the total number of quarks to be greater than $D_\mathrm{quality}$ while keeping $N_\mathrm{DW}=1$.} The theory is thus constrained from two sides. The axion quality problem favors big axions coming from high-dimension network Higgs monomials, but the requirement that QCD remains asymptotically free limits how high-dimension those monomials can be. Any theory which contains a high-quality QCD axion whilst simultaneously maintaining the asymptotic freedom of QCD must therefore be a \textit{little} big QCD axion.

\section{\label{sec:the-littlest-big-axion}Little Big QCD Axions}

We now construct a phenomenologically viable little big QCD axion model which provides a high quality solution to the strong CP problem while keeping QCD  asymptotically free and remaining cosmologically consistent. There is no unique theory space capable of achieving this. In the following, we build on the star big axion theory space introduced in Sec.~\ref{sec:big-axions} with charge matrix~\eqref{eq:star} and demonstrate how it can realize the QCD axion of our universe.

To make the star big axion phenomenologically viable, we take $q=5$ and endow it with the fermion content 
\begin{equation}
    \mathbf{d} = \begin{pmatrix}
        3 \\
        0 \\ 
        3 \\ 
        0
    \end{pmatrix}\, , \hspace{10mm} \mathbf{k} =     \begin{pmatrix}
        2 \\
        1 \\ 
        2 \\ 
        5
    \end{pmatrix}\, .
\end{equation}
This model is protected against quality-violating operators up to operator dimension $D_\mathrm{quality}=16$, ensuring there is no quality problem provided $f_\mathrm{geo}=(f_1^5 f_2 f_3^5 f_4^5)^{1/16}\lesssim 10^{13}\,\mathrm{GeV}$. This corresponds to an allowed effective decay constant of $f_\mathrm{eff}\lesssim 10^{12}\,\mathrm{GeV}$ assuming that the individual $f_i$'s are within a few orders of magnitude of each other.\footnote{Quality directly constrains $f_\mathrm{geo}$ rather than $f_\mathrm{eff}$. For the star big axion, taking all decay constants equal yields $f_\mathrm{geo}=\sqrt{76} f_\mathrm{eff}$, however large hierarchies between the $f_i$ can drive down $f_\mathrm{eff}$ further relative to $f_\mathrm{geo}$.}  The theory has domain wall number $N_\mathrm{DW}=1$ and contains only $10$ new heavy quarks, sufficiently few that QCD remains asymptotically free.

Beyond these basic considerations, the model has additional $U(1)_\alpha\times U(1)_\beta \times U(1)_Y$ and $U(1)_\alpha\times U(1)_\beta \times U(1)_\gamma$ gauge anomalies that must be canceled for the theory to be mathematically self-consistent. The additional matter content required to achieve these cancellations depends on whether the model needs to be consistent with pre- or post-inflationary cosmology. This is worked out in detail in Appendix~\ref{app:abelian-anomaly-counting}, but the summary is as follows.

If the big axion is pre-inflationary, there is virtually no restriction on the exact charge assignment for the extra anomaly-canceling fermions. The ones that carry hypercharge will contribute to the running of $\alpha_Y$, but since the fermion masses are naturally $\mathcal{O}(f_i)$, the associated Landau pole can be kept well above the Planck scale. If the cosmology is post-inflationary and the fermion masses are below the reheating scale of the universe, the hypercharge assignments are less generic. As pointed out by Refs.~\cite{DiLuzio:2016sbl,DiLuzio:2017pfr,Lu:2023ayc}, if KSVZ fermions are neutral under both $SU(2)_L$ and $U(1)_Y$, they can combine with SM quarks to form stable fractionally charged hadrons, the existence of which is strongly constrained by experimental observations. In our case, an appropriate choice of charge assignment is to let the heavy right-handed KSVZ quarks be down-like, i.e. have hypercharge $\pm 1/3$ and be neutral under $SU(2)_L$. The additional $U(1)^3$ and $U(1)\times(\mathrm{gravity})^2$ anomaly cancellations can then be achieved with fermions that either have integer hypercharge, allowing them to decay to SM leptons, or are completely neutral under the SM. With this setup, the hypercharge Landau pole is above the Planck scale, and the ratio of the electromagnetic and color anomalies in the star big axion theory is  $E/N=2/3$. Different values for $E/N$, which contributes to the effective axion-photon coupling, can be attained by changing the charge assignments however.

An additional feature of the post-inflationary scenario is the presence of axion strings and domain walls in the early universe~\cite{Sikivie:1982qv,Sikivie:2006ni}. As pointed out by Ref.~\cite{Lu:2023ayc}, when the axion is a linear combination of the phases of multiple complex scalars, the number of domain walls attached to each string need not coincide with $N_\mathrm{DW}$. Instead, if the field $\Phi_i$ enters into the big axion with multiplicity $|n_i|>1$, the strings that form at $T\sim f_i$ will have effective domain wall number $N_\mathrm{DW}|n_i|$, potentially leading to a domain wall problem. Big axions can avoid these potential issues provided at least one of the network Higgs fields has multiplicity one. In that case, the theory contains a one-wall string candidate. Whether this string is populated in a general post-inflationary history, and how it interacts with the other gauged/global strings in the network, is a dynamical question that we leave for future work.\footnote{See Refs.~\cite{Hiramatsu:2019tua,Hiramatsu:2020zlp,Correia:2022spe,Niu:2023khv,Lee:2024toz,Mupo:2025ner} for a detailed discussion of multi-string dynamics in multi-axion and Abelian Higgs models, of relevance to the most general post-inflationary big axion cosmology.} However, in specific cosmologies where the multiplicity-one network Higgs field also has the lowest symmetry restoration temperature and the reheating temperature exceeds this scale but remains below those associated with Higgs fields of higher multiplicity, only strings with an effective domain wall number of one are formed, and the axion string-wall network is able to collapse.

Little big axions constitute compelling QCD axion candidates in both pre- and post-inflationary scenarios. If explicit PQ-breaking operators are absent through dimension $15$ as in the example outlined above, sufficiently high axion quality can be maintained for effective axion decay constants as large as $f_{\rm eff}\lesssim 10^{12}\,\mathrm{GeV}$. This places these models in a regime in which the QCD axion could viably comprise some or all of the DM abundance in suitable pre- or post-inflationary histories. Other big axion models may exhibit even greater quality protection, allowing for still larger decay constants. It is evident that the model building possibilities afforded by the big axion framework are vast and remain largely open.

\section{Outlook}\label{sec:outlook}
In this work, we have introduced \textit{big axions}: a general class of axion models in which the axion emerges as a light Nambu--Goldstone mode associated with the collective spontaneous breaking of a set of $U(1)$ symmetries within a theory space consisting of linked Higgs and Abelian gauge fields. This framework unifies several previously studied constructions while revealing a much broader landscape of protected axion models. The defining feature of this framework is that the quality of the axion is controlled by the global structure of the network: any operator capable of explicitly breaking the axion shift symmetry necessarily probes the entire theory space, and is therefore  pushed to high  dimension. With the addition of appropriate fermion content, we have demonstrated that big axions can be consistently coupled to QCD to solve the strong CP problem with high quality. We further identified a minimal phenomenologically viable subclass of \textit{little big QCD axions} which preserve the asymptotic freedom of QCD, can accommodate both pre-inflationary cosmologies and suitable post-inflationary histories, and may constitute the DM of our universe.

One of the most striking features of big axions is that highly protected accidental symmetries can arise even in purely four-dimensional gauge theories from comparatively simple ingredients. The dual moose formulation reveals that, upon moving beyond nearest-neighbor linking, theory space can admit a rich variety of topological structures. These topologies govern the kernel of the charge matrix and ultimately determine the existence and quality of the emergent Nambu--Goldstone modes. The examples studied in this work likely represent only a small corner of a much broader landscape whose full phenomenological and theoretical implications remain to be uncovered. In particular, more general charge matrices, potentially requiring hypergraph-like descriptions, may reveal new classes of accidental symmetries and protected light fields. The connection between big axions and deconstruction also hints at a deeper interplay between theory-space topology, homology, higher-form gauge fields, and generalized symmetries which warrants further investigation.

The cosmology of big axions also merits further study. Since the big axion is realized as a collective mode built from the phases of multiple Higgs fields, the resulting string and domain wall networks in post-inflationary cosmologies can differ substantially from  single field axion models. In particular, different network Higgs fields may form strings with different effective domain wall numbers, leading to a mixed gauge-global string network whose evolution is sensitive to both the hierarchy of symmetry breaking scales and the reheating temperature. Understanding these dynamics is essential for delineating the viable parameter space of little big QCD axions, refining their predictions as dark matter candidates, and determining whether they give rise to additional observable signatures, such as gravitational waves. We have demonstrated that, with appropriate charge assignments, the star little big QCD axion is compatible with a post-inflationary cosmological history. To the best of our knowledge, this is the first explicit weakly coupled four-dimensional high-quality QCD axion construction that simultaneously addresses the known post-inflationary obstacles: $N_{\rm DW}=1$, the existence of a one-wall string candidate, controlled colored matter, absence of stable fractionally charged relics, and perturbative SM running up to the Planck scale. This places the big axion framework in a rare corner of the axion model landscape, populated by only a handful of extra-dimensional models~\cite{Lu:2023ayc,Petrossian-Byrne:2025mto,Loladze:2025uvf}, in which high quality PQ symmetry protection and post-inflationary viability are simultaneously realized, rendering the study of its distinctive cosmological dynamics particularly intriguing.

The applications of big axions may extend well beyond the strong CP problem and DM. By furnishing a simple yet robust mechanism for generating high-quality accidental symmetries in four dimensional theories, the framework opens new directions for inflation, quintessence, relaxion dynamics, and other phenomena which rely on protected scalar sectors. It is likewise intriguing to ask whether analogous constructions could exist for non-Abelian theories, further broadening the possible use cases, in addition to whether the framework discussed could offer a simultaneous explanation of the SM flavor hierarchies. 

Taken together, the results of this work suggest that topology and non-locality in theory space may provide a powerful organizing principle for the generation of protected low-energy structure, with rich implications for axion physics, cosmology, and the broader understanding of emergent symmetries.

\begin{acknowledgments}

We thank Qianshu Lu for helpful discussions, and Nathaniel Craig, David E. Kaplan, Anson Hook, Qianshu Lu, Mario Reig, and Martin Schmaltz for useful comments on the manuscript. H.B.\ is supported by the James Arthur Postdoctoral Fellowship at New York University. M.K.\ is supported by a James Arthur Graduate Assistantship at New York University. N.W.\ is supported by the NSF under grant PHY-2210498, by the BSF under grant 2018140, and by the Simons Foundation. We acknowledge the use of generative AI tools, including \texttt{GPT 5.5} and \texttt{Claude Opus 4.7}, for technical assistance in editing, literature search, and matrix and anomaly combinatorics, all of which were human verified.
\end{acknowledgments}
\setlength{\textfloatsep}{8pt plus 2pt minus 2pt}
\setlength{\floatsep}{6pt plus 2pt minus 2pt}
\setlength{\intextsep}{8pt plus 2pt minus 2pt}
\setlength{\abovecaptionskip}{4pt}
\setlength{\belowcaptionskip}{0pt}

\appendix

\vspace{-0.30cm}
\section{General Big Axions}
\label{app:general}
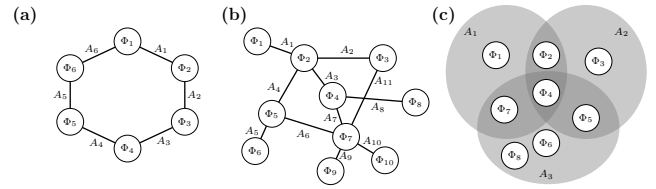
\begin{figure}
    \centering
    \resizebox{\columnwidth}{!}{%
\begin{tikzpicture}[
    site/.style={
        circle,
        draw,
        fill=white,
        minimum size=6.4mm,
        inner sep=0pt,
        font=\scriptsize
    },
    link/.style={
        line width=0.9pt
    },
    panellab/.style={
        font=\bfseries\large
    },
    glabel/.style={
        font=\scriptsize,
        inner sep=0.8pt
    },
    regionlabel/.style={
        font=\scriptsize,
        inner sep=1pt
    }
]

\def\sep{5.00}

\def\panelxmin{-2.75}
\def\panelxmax{ 2.75}
\def\panelymin{-2.2}
\def\panelymax{ 2.2}

\begin{scope}[shift={(0,0)}]
    \path[use as bounding box]
        (\panelxmin,\panelymin) rectangle (\panelxmax,\panelymax);

    \node[panellab] at (-2.45,1.90) {(a)};

    \foreach \i/\ang in {
        1/90,
        2/30,
        3/-30,
        4/-90,
        5/-150,
        6/150
    }{
        \coordinate (C\i) at ({1.58*cos(\ang)},{1.28*sin(\ang)});
    }

    \draw[link] (C1)--(C2)--(C3)--(C4)--(C5)--(C6)--(C1);

    \node[glabel] at ( 0.8, 1.1) {$A_1$};
    \node[glabel] at ( 1.6, 0.00) {$A_2$};
    \node[glabel] at ( 0.87,-1.10) {$A_3$};
    \node[glabel] at (-0.74,-1.14) {$A_4$};
    \node[glabel] at (-1.60,0.00) {$A_5$};
    \node[glabel] at (-0.85, 1.1) {$A_6$};

    \node[site] at (C1) {$\Phi_1$};
    \node[site] at (C2) {$\Phi_2$};
    \node[site] at (C3) {$\Phi_3$};
    \node[site] at (C4) {$\Phi_4$};
    \node[site] at (C5) {$\Phi_5$};
    \node[site] at (C6) {$\Phi_6$};
\end{scope}

\begin{scope}[shift={(\sep,0)}]
    \path[use as bounding box]
        (\panelxmin,\panelymin) rectangle (\panelxmax,\panelymax);

    \node[panellab] at (-2.45,1.90) {(b)};

    \coordinate (B1)  at (-1.90, 1.28);
    \coordinate (B2)  at (-0.78, 0.88);
    \coordinate (B3)  at ( 1.10, 0.88);
    \coordinate (B4)  at (-0.10,-0.02);
    \coordinate (B5)  at (-1.55,-0.45);
    \coordinate (B6)  at (-1.95,-1.34);
    \coordinate (B7)  at ( 0.22,-1.00);
    \coordinate (B8)  at ( 1.88,-0.18);
    \coordinate (B9)  at (-0.15,-1.82);
    \coordinate (B10) at ( 1.15,-1.56);

    \draw[link] (B1) -- (B2);   
    \draw[link] (B2) -- (B3);   
    \draw[link] (B2) -- (B4);   
    \draw[link] (B2) -- (B5);   
    \draw[link] (B5) -- (B6);   
    \draw[link] (B5) -- (B7);   
    \draw[link] (B4) -- (B7);   
    \draw[link] (B4) -- (B8);   
    \draw[link] (B7) -- (B9);   
    \draw[link] (B7) -- (B10);  
    \draw[link] (B3) -- (B7);   

    \node[glabel] at (-1.18, 1.25) {$A_1$};
    \node[glabel] at ( 0.25, 1.08) {$A_2$};
    \node[glabel] at (-0.12, 0.46) {$A_3$};
    \node[glabel] at (-1.50, 0.18) {$A_4$};
    \node[glabel] at (-2.02,-0.86) {$A_5$};
    \node[glabel] at (-0.80,-0.93) {$A_6$};
    \node[glabel] at ( -0.15,-0.55) {$A_7$};
    \node[glabel] at ( 0.98,-0.30) {$A_8$};
    \node[glabel] at (0.21,-1.46) {$A_9$};
    \node[glabel] at ( 0.85,-1.10) {$A_{10}$};
    \node[glabel] at ( 1.12, 0.34) {$A_{11}$};

    \node[site] at (B1)  {$\Phi_1$};
    \node[site] at (B2)  {$\Phi_2$};
    \node[site] at (B3)  {$\Phi_3$};
    \node[site] at (B4)  {$\Phi_4$};
    \node[site] at (B5)  {$\Phi_5$};
    \node[site] at (B6)  {$\Phi_6$};
    \node[site] at (B7)  {$\Phi_7$};
    \node[site] at (B8)  {$\Phi_8$};
    \node[site] at (B9)  {$\Phi_9$};
    \node[site] at (B10) {$\Phi_{10}$};
\end{scope}

\begin{scope}[shift={(2*\sep,0)}]
    \path[use as bounding box]
        (\panelxmin,\panelymin) rectangle (\panelxmax,\panelymax);

    \node[panellab] at (-2.45,1.90) {(c)};

    \fill[
        black!55,
        fill opacity=0.38,
        draw=none
    ] (-0.95,0.55) ellipse [x radius=1.45, y radius=1.60];

    \fill[
        black!55,
        fill opacity=0.38,
        draw=none
    ] (0.95,0.55) ellipse [x radius=1.45, y radius=1.60];

    \fill[
        black!55,
        fill opacity=0.38,
        draw=none
    ] (0.05,-0.78) ellipse [x radius=1.70, y radius=1.35];

    \node[regionlabel] at (-1.80,1.50) {$A_1$};
    \node[regionlabel] at ( 1.80,1.50) {$A_2$};
    \node[regionlabel] at ( 0.00,-1.9) {$A_3$};

    \coordinate (H1) at (-1.20, 0.95);
    \coordinate (H2) at ( 0.00, 0.95);
    \coordinate (H3) at ( 1.25, 0.80);
    \coordinate (H4) at (0.00, 0.05);
    \coordinate (H5) at ( 0.95,-0.55);
    \coordinate (H6) at ( 0.00,-1.15);
    \coordinate (H7) at (-1.0,-0.35);
    \coordinate (H8) at (-0.75,-1.45);

    \node[site] at (H1) {$\Phi_1$};
    \node[site] at (H2) {$\Phi_2$};
    \node[site] at (H3) {$\Phi_3$};
    \node[site] at (H4) {$\Phi_4$};
    \node[site] at (H5) {$\Phi_5$};
    \node[site] at (H6) {$\Phi_6$};
    \node[site] at (H7) {$\Phi_7$};
    \node[site] at (H8) {$\Phi_8$};
\end{scope}

\end{tikzpicture}
}
\caption{Additional examples of big axion theory spaces. (a) A geometric circular big axion theory space of the kind discussed in Ref.~\cite{Hill:2002kq}. (b) A random tree-like geometric big axion theory space. (c) An accidental hypergraph big axion theory space. With an appropriate choice of charge assignments, all of these theory space structures admit Nambu--Goldstone bosons with various degrees of quality protection.}
    \label{fig:extra-diagrams}
\end{figure}

In Sec.~\ref{sec:big-axions}, we introduced the tetrahedral and star big axions, both of which admit a \textit{dual moose} description. The big axion framework encompasses a much broader class of models however, whose associated theory spaces are most generally described by \textit{hypergraphs} rather than ordinary graphs. To illustrate the diversity of this class of models, we provide some additional examples of big axion theory spaces in this Appendix. The graphs of these theories are shown in Fig.~\ref{fig:extra-diagrams}.
The first, labeled (a), is a circular dual moose. A suitable choice of charge matrix for this graph is
\begin{equation}
    \hat{\mathbf{Q}}_1 = \begin{pmatrix}
        1 & -1 & 0 & 0 & 0 & 0 \\
        0 & 1 & -1 & 0 & 0 & 0 \\
        0 & 0 & 1 & -1 & 0 & 0 \\
        0 & 0 & 0 & 1 & -1 & 0 \\
        0 & 0 & 0 & 0 & 1 & -1 \\
        -1 & 0 & 0 & 0 & 0 & 1
    \end{pmatrix} ,
\end{equation}
which has a single big axion $\mathbf{n}^T=(1,1,1,1,1,1)$, corresponding to a deconstructed $A_5$ zero-mode from a circular extra dimension~\cite{Hill:2002kq}. The second theory space, labeled (b), is also a dual moose and can be associated with the charge matrix
\begin{equation}
    \hat{\mathbf{Q}}_2 = \begin{pmatrix}
        1 & -3 & 0 & 0 & 0 & 0 & 0 & 0 & 0 & 0 \\
        0 & 1 & 3 & 0 & 0 & 0 & 0 & 0 & 0 & 0 \\
        0 & -2 & 0 & 4 & 0 & 0 & 0 & 0 & 0 & 0 \\
        0 & 1 & 0 & 0 & 1 & 0 & 0 & 0 & 0 & 0 \\
        0 & 0 & 0 & 0 & -2 & -1 & 0 & 0 & 0 & 0 \\
        0 & 0 & 0 & 0 & -1 & 0 & 1 & 0 & 0 & 0 \\
        0 & 0 & 0 & 6 & 0 & 0 & 3 & 0 & 0 & 0 \\
        0 & 0 & 0 & -1 & 0 & 0 & 0 & -1 & 0 & 0 \\
        0 & 0 & 0 & 0 & 0 & 0 & -1 & 0 & 1 & 0 \\
        0 & 0 & 0 & 0 & 0 & 0 & 2 & 0 & -2 & 0 \\
        0 & 0 & 0 & 0 & 0 & 0 & 1 & 0 & 0 & -1

    \end{pmatrix}\,.
\end{equation}
This also has a single big axion $\mathbf{n}^T=(-18,-6,2,-3,6,-12,6,3,6,6)$ which is notably protected from PQ symmetry breaking up to operator dimension $D_\mathrm{quality}=68$. Finally, a choice of charge matrix to go with the third \textit{hypergraph} (c) is
\begin{equation}
    \hat{\mathbf{Q}}_3 = \begin{pmatrix}
        1 & 2 & 0 & -1 & 0 & 0 & 1 & 0 \\
        0 & -1 & 2 & 1 & -1 & 0 & 0 & 0 \\
        0 & 0 & 0 & 0 & 1 & -1 & -2 & 1 
    \end{pmatrix},
\end{equation}
which has five big axions $\mathbf{n}^T_1=(-2,1,0,0,-1,0,0,1)$, $\mathbf{n}^T_2=(3,-2,0,0,2,0,1,0)$, $\mathbf{n}^T_3=(2,-1,0,0,1,1,0,0)$, $\mathbf{n}^T_4=(-1,1,0,1,0,0,0,0)$, and $\mathbf{n}^T_5=(-4,2,1,0,0,0,0,0)$. While we refrain from working it out in detail here, like the examples addressed in the body of this Letter, these theories can be coupled to QCD by introducing KSVZ- and DFSZ- like couplings to SM and BSM fermions, with any remaining $U(1)^3$ and $U(1)\times(\mathrm{gravity})^2$ gauge anomalies canceled by an additional population of fermionic matter with appropriate charge assignments. For these big axions to simultaneously provide a controlled solution to the quality problem, the Landau poles associated with QCD and $U(1)_Y$ would have to reside above $M_\mathrm{Pl}$. These examples demonstrate that, while not guaranteed, big axions generically arise in a large subset of theory spaces.

\vspace{-0.65\baselineskip}
\section{QCD Anomaly Counting}\label{app:qcd-anomaly-counting}
In this Appendix, we derive Eq.~\eqref{eq:qcd-anomaly-cancelation-counting}; a consistency condition on fermion couplings in the big axion network in order for the big axion theory to be free of $U(1)_\alpha\times SU(3)_c\times SU(3)_c$ gauge anomalies. When adding Yukawa couplings of the form $\Phi_i \psi_A \psi_A^c$ to the big axion theory, where $\Phi_i$ are network Higgs fields and $\psi_A$ and $\psi^c_A$ are generic fermions in the $\mathbf{3}$ and $\bar{\mathbf{3}}$ of $SU(3)_c$, respectively, the gauge anomaly is captured by the coefficient
\begin{equation}\label{eq:anomaly-coefficient}
    \mathcal{A}_\alpha = \frac{1}{2}\sum_{A}\left(q_\alpha[\psi_A] + q_\alpha[\psi_A^c]\right) \, ,
\end{equation}
where $q_\alpha[\psi_A]$ is the charge of the $\psi_A$ field under the $U(1)_\alpha$ gauge group. This coefficient must vanish for the theory to be gauge invariant. We now show that setting Eq.~\eqref{eq:anomaly-coefficient} to zero is equivalent to requiring that the determinant of the quark mass matrix in the big axion network is gauge invariant, and use this to deduce Eq.~\eqref{eq:qcd-anomaly-cancelation-counting}.

For a KSVZ-like coupling, given by Eq.~\eqref{eq:ksvz-coupling}, gauge invariance requires that
\begin{equation}
    q_\alpha[\Psi_A] + q_\alpha[\Psi^c_A]=\hat{Q}_{\alpha i_A} \, ,
\end{equation}
where the index $i_A$ labels the network Higgs field coupled to $\Psi_A$. Let the vector $\mathbf{k}\in\mathbb{Z}^{N_\Phi}$ indicate the multiplicity of KSVZ-like quarks coupled to the network Higgs fields via couplings of the form~\eqref{eq:ksvz-coupling}. The  determinant of the quark mass matrix is then
\begin{equation}\label{eq:ksvz-determinant}
    \mathrm{det}\left(M_\Psi\right) \propto \prod_{j=1}^{N_\Phi} (\Phi_j^*)^{k_j} \, .
\end{equation}
Under the network of $U(1)_\alpha$ gauge transformations, Eq.~\eqref{eq:ksvz-determinant} transforms as
\begin{equation}\label{eq:ksvz-det-transform}
    \delta\,\mathrm{arg}\,\mathrm{det} \left(M_\Psi\right) = -\sum_\alpha\lambda_\alpha (\hat{\mathbf{Q}}\mathbf{k})_\alpha \, ,
\end{equation}
where $\lambda_\alpha \in [0,2\pi)$ is a $U(1)_\alpha$ gauge transformation parameter. Notice that if the gauge anomaly comes only from KSVZ-type couplings, Eq.~\eqref{eq:anomaly-coefficient} vanishes if and only if Eq.~\eqref{eq:ksvz-det-transform} is trivial.\footnote{In the most general case where hidden $U(1)_\alpha$ charge is assigned to both $\Psi_A$ and $\Psi_A^c$ rather than just one of them, the determinant may be a sum of monomials in the $\Phi_i$'s and in masses. For a general mass matrix $\mathbf{M}$ arising from terms of the form $\Psi_A M_{AB} \Psi_B^c + \mathrm{h.c.}$, the relation~\eqref{eq:anomaly-gauge-cancelation-condition} still holds, but one would have to ensure that additional corrections to the effective axion potential  are not induced (i.e. that there is a genuine global $U(1)_\mathrm{PQ}$ symmetry in the theory in the first place).}

For a DFSZ-type coupling, given by Eq.~\eqref{eq:dfsz-coupling}, gauge invariance requires
\begin{align}
    q_\alpha[Q^a] + q_\alpha[(u^c)^a]&=-q_\alpha[H_u] \, ,\\ q_\alpha[Q^a] + q_\alpha[(d^c)^a]&=-q_\alpha[H_d] \, ,
\end{align}
and
\begin{equation}\label{eq:higgs-to-pq-charges}
    \left(q_\alpha[H_u] + q_\alpha[H_d]\right) = -\left(Q_{\alpha i} + Q_{\alpha j}\right)\, ,
\end{equation}
where $i$ and $j$ are indicating that network Higgs fields $\Phi_i$ and $\Phi_j$ enter into the coupling~\eqref{eq:dfsz-coupling}, $\alpha$ indexes the gauge groups $U(1)_\alpha$, and $a$ labels the flavor generation. Let the vector $\mathbf{d}\in\mathbb{Z}^{N_\Phi}$ indicate the multiplicity of the SM quarks that couple to the network Higgs fields  via terms of the form~\eqref{eq:dfsz-coupling}. Unlike $\mathbf{k}$, unless additional flavor structure is added, $\mathbf{d}$ will only have one or two non-zero entries taking values of $\pm 6$ or $\pm 3$ respectively, where the factor of three arises from the three SM generations. Using Eq.~\eqref{eq:higgs-to-pq-charges}, the mass matrix for the DFSZ-coupled quarks transforms as
\begin{equation}\label{eq:dfsz-det-transform}
    \delta\,\mathrm{arg}\,\det (M_u) \det (M_d) = -\sum_\alpha \lambda_\alpha (\hat{\mathbf{Q}}\mathbf{d})_\alpha \, .
\end{equation}
Notice that if the gauge anomaly comes only from DFSZ-type couplings, Eq.~\eqref{eq:dfsz-det-transform} is trivial exactly when Eq.~\eqref{eq:anomaly-coefficient} vanishes.

For the \textit{total} quark mass matrix determinant to be gauge invariant, including both KSVZ- and DFSZ-type quark couplings, we thus need
\begin{equation}\label{eq:anomaly-gauge-cancelation-condition}
    \hat{\mathbf{Q}}\left(\mathbf{k}+\mathbf{d}\right) = 0 \, ,
\end{equation}
which, due to the charge relationships, is equivalent to Eq.~\eqref{eq:anomaly-coefficient} vanishing for all $U(1)_\alpha$ gauge groups. Recall that $\mathbf{n}$ is defined to be a primitive vector satisfying $\hat{\mathbf{Q}}\mathbf{n}=0$. Hence, for a single big gauge anomaly-free axion
\begin{equation}
    \mathbf{k}+\mathbf{d} = N_\mathrm{DW} \mathbf{n}\, ,
\end{equation}
with $N_\mathrm{DW}\in \mathbb{Z}$ being the axion domain wall number. Stated differently, the gauge invariance of the quark mass matrix determinant forces its phase to be exactly proportional to $\Theta_\mathbf{n}$ with proportionality constant $N_\mathrm{DW}$. Removing this phase via fermion field redefinitions then gives rise to an effective axion potential of the form
\begin{equation}\label{eq:effective-qcd-axion-potential}
    V_{\mathrm{QCD}}\propto
    \cos(N_{\rm DW}\Theta_{\mathbf n}+\bar\theta)\, ,
\end{equation}
due to the non-invariance of the path integral measure, with $\bar{\theta}$ being the axion independent part of the phase.

\vspace{-0.65\baselineskip}
\section{Fractionally Charged Relics and Hypercharge Anomaly Counting}\label{app:abelian-anomaly-counting}

As pointed out in Refs.~\cite{DiLuzio:2016sbl,DiLuzio:2017pfr,Lu:2023ayc}, the presence of additional colored fields can be cosmologically problematic. If the universe reheats at a sufficiently high temperature to produce these particles, they can bind with SM quarks to produce color-neutral bound states with fractional electric charge. An obvious resolution to this is to also endow these new KSVZ-like quarks with hypercharge.

Given the large number of gauge groups already present in the theory, this makes the question of anomaly cancellation non-trivial. In addition, as pointed out in Ref.~\cite{Lu:2023ayc}, the presence of additional hypercharged matter modifies the running of the SM gauge couplings, hastening the appearance of a Landau pole. This constraint can be formulated in a reasonably model-independent fashion as follows.

Taking the individual fermion masses to lie at approximately the geometric mean of the decay constants that enter into the big axion, $f_\mathrm{geo}=(f_1^5 f_2 f_3^5 f_4^5)^{1/16}$, the budget on $\Delta b_{Y}$, the shift in the one-loop hypercharge beta function coefficient, is
\begin{equation}
    \Delta b_{Y}^{\rm budget}
    \;=\;
    \begin{cases}
        21.1 & f_{\rm geo} = 10^{12}~\text{GeV}\, , \\[2pt]
        13.4 & f_{\rm geo} = 10^{8}~\text{GeV}\, .
    \end{cases}
    \label{eq:Ybudgets}
\end{equation}
This shows that the hypercharge $\beta$-function coefficient can grow by about
$21$ above its SM value $b_{Y}^{\rm SM}=41/6\simeq 6.83$ before the
$U(1)_{Y}$ Landau pole drops below $\Mpl$ when $f_{\rm geo} = 10^{12}$~GeV,
and by about $13$ when $f_{\rm geo} = 10^{8}$~GeV.\footnote{To be precise, the scale dictating the hypercharge running is  the geometric mean of the individual hypercharged quark mass scales. For example, for the KSVZ sector, this scale is $f_Y = (f_1^2 f_2 f_3^2 f_4^5)^{1/10}$, coinciding with $f_{\rm geo}$ if all the $f_i$ are comparable. By introducing large hierarchies among the $f_i$ however,  $f_Y$ can be raised relative to $f_{\rm geo}$,  delaying the onset of additional hypercharge running and further alleviating Landau-pole constraints. Since this is unnecessary for the benchmark model considered here, we do not pursue this possibility.} 

The total contribution from the $|\mathbf k|$
new pairs of quarks added at the scale $f_{\rm geo}$ is simply
\begin{equation}
\label{eq:DeltabY-from-sumY2}
    \Delta b^K_{Y} \;=\; 4\,\sum_{\rm pairs}Y^{2}.
\end{equation}
Meanwhile, the second Higgs doublet, which we take to lie at the electroweak scale, contributes an effective
\begin{equation}
    \Delta b_Y^H = \frac{1}{6} \frac{\log{M_\mathrm{Pl}/M_\mathrm{H}}}{\log{M_\mathrm{Pl}/f_\mathrm{geo}}}=\begin{cases}
        0.4&  f_\mathrm{geo}=10^{12}\,\mathrm{GeV}\, ,\\
        0.3& f_\mathrm{geo}=10^{8}\,\mathrm{GeV}\, ,
    \end{cases}
\end{equation}
at the scale $f_\mathrm{geo}$. This leads to limits of $\Delta b^{K}_{Y}<20.7$ ($13.1$) for $f_{\rm geo}=10^{12}\gev$ ($f_{\rm geo}=10^8 \gev$). We focus on a single `all down-type' charge assignment  for which all the KSVZ-like quarks carry hypercharge $\pm \frac{1}{3}$. Specific charge values are shown in
Table~\ref{tb:star-charges}. With this assignment, the KSVZ-like quark contribution to the running is $\Delta b_Y^K = 0.44 |\mathbf{k}|=4.44$. Thus, accounting for the SM matter, the extra Higgs doublet, and the new KSVZ quarks alone, the hypercharge Landau pole naturally lies far above the Planck scale. However, we must also address any remaining $U(1)^3$ and $U(1)\times(\mathrm{gravity})^2$ gauge anomalies, which need to be canceled by additional charged matter.

\begin{table}[t]
\centering
\renewcommand{\arraystretch}{1.15}
\begin{tabular}{l|cccc}
\hline\hline
                  & \multicolumn{4}{c}{down-type ($Y_\Psi=-\tfrac{1}{3}$)}\\
field             & $A_1$ & $A_2$ & $A_3$ & $Y$\\
\hline
$\Phi_1$          & $-1$  & $-1$  & $-1$  & $0$\\
$\Phi_2$          & $5$   & $0$   & $0$   & $0$\\
$\Phi_3$          & $0$   & $1$   & $0$   & $0$\\
$\Phi_4$          & $0$   & $0$   & $1$   & $0$\\
\hline
$H_u$             & $1$   & $0$   & $1$   & $+\tfrac{1}{2}$\\
$H_d$             & $0$   & $0$   & $0$   & $-\tfrac{1}{2}$\\
$Q^a$             & $0$   & $0$   & $0$   & $+\tfrac{1}{6}$\\
$L^a$             & $0$   & $0$   & $0$   & $-\tfrac{1}{2}$\\
$(u^{a})^{c}$         & $-1$  & $0$   & $-1$  & $-\tfrac{2}{3}$\\
$(d^{a})^{c}$         & $0$   & $0$   & $0$   & $+\tfrac{1}{3}$\\
$(e^{a})^{c}$         & $0$   & $0$   & $0$   & $+1$\\
\hline
$\Psi_A^{(1)}$ ($k_1{=}2$) & $-1$ & $-1$ & $-1$ & $-\tfrac{1}{3}$\\
$(\Psi_A^{(1)})^{c}$           & $0$  & $0$  & $0$  & $+\tfrac{1}{3}$\\
$\Psi^{(2)}$ ($k_2{=}1$)       & $5$  & $0$  & $0$  & $-\tfrac{1}{3}$\\
$(\Psi^{(2)})^{c}$               & $0$  & $0$  & $0$  & $+\tfrac{1}{3}$\\
$\Psi_A^{(3)}$ ($k_3{=}2$) & $0$  & $1$  & $0$  & $-\tfrac{1}{3}$\\
$(\Psi_A^{(3)})^{c}$           & $0$  & $0$  & $0$  & $+\tfrac{1}{3}$\\
$\Psi_A^{(4)}$ ($k_4{=}5$) & $0$  & $0$  & $1$  & $-\tfrac{1}{3}$\\
$(\Psi_A^{(4)})^{c}$           & $0$  & $0$  & $0$  & $+\tfrac{1}{3}$\\
\hline\hline
\end{tabular}
\caption{Hidden $U(1)_\alpha$ and hypercharge assignments for the star big axion model with $\mathbf{k}=(2,1,2,5)$ and $\mathbf{d}=(3,0,3,0)$. KSVZ quark Dirac masses arise from terms $\lambda_A\,\Phi_{i_A}^{\ast}\Psi_A^{(i_A)}(\Psi_A^{(i_A)})^{c}$ such that $q_\alpha[\Psi_A]+q_\alpha[\Psi_A^c]=+Q_{\alpha,i_A}$. Here, the bracketed superscript on the fermions indicates the Higgs field to which the fermion is coupled, and the subscript $A$ is over all fermions coupled at that site. We assume a hidden charge site-localized split: $q_\alpha[\Psi_A]=+Q_{\alpha,i_A}$, $q_\alpha[\Psi_A^c]=0$.}
\label{tb:star-charges}
\end{table}

In the presence of the matter shown in Table~\ref{tb:star-charges}, several anomaly classes vanish by construction. The $U(1)_\alpha\times SU(3)_c^2$ anomalies cancel because the KSVZ and DFSZ quark multiplicities  satisfy $\mathbf{k}+\mathbf{d}=\mathbf{n}\in\ker\hat{\mathbf{Q}}_{\rm star}$, leaving exactly the chiral $G\tilde G$ coupling along the big axion direction that solves the strong CP problem (Eq.~\eqref{eq:effective-qcd-axion-potential}).  Every KSVZ-like quark contribution to a $U(1)_Y$ anomaly, specifically $U(1)_Y\times SU(3)_c^2$, $U(1)_Y^3$ and  $U(1)_Y\times(\mathrm{gravity})^2$, cancels pair-by-pair because each $(\Psi_A,\Psi_A^c)$ has $Y_\Psi+Y_{\Psi^c}=0$, leaving only the (anomaly-free) SM contributions. A further set of cancellations happens along the $A_2$ direction: because $\Psi_1$ and $\Psi_3$ have $q_2=\pm 1$ with the same multiplicity ($k_1=k_3=2$) and the DFSZ-coupled sites are similarly symmetric ($d_1=d_3$), the following triangle anomalies carrying an \emph{odd} power of $q_2$ vanish automatically: $A_2\times SU(3)_c^2$, $A_2\times U(1)_Y^2$, $A_2\times(\mathrm{gravity})^2$, and $A_2^3$. 

Beyond this there remain a number of anomalies, including mixed hypercharge-network charge anomalies, and cubic anomalies, which  can all be addressed without any additional colored matter. As an example, we provide a set of colorless fields which can cancel these anomalies in Table \ref{tb:star-cancellers}. These fields all have hypercharge $0$ or $\pm 1$ and thus have electric charge $0$ or $\pm 1$. The presence of these fields contributes  $\Delta b_Y^\chi=16$ to the hypercharge beta function to yield a total contribution $\Delta b_Y^{K+\chi}=20.44$. This lies within the one-loop $f_{\rm geo}= 10^{12}\,\mathrm{GeV}$ budget, such that the effective theory remains valid up to the Planck scale. It does not lie within the $f_\mathrm{geo}=10^{8} \,\mathrm{GeV}$ budget, disfavoring this scenario unless $U(1)_Y$ can be embedded into a suitable, controlled UV completion. Since this construction involves the addition of a large number of additional states, it is also subject to the species bound, in which the effective gravitational cut-off is lowered to $\Lambda_{\rm sp} = M_{\mathrm{Pl}}/\sqrt{N}$, where $N$ denotes the total number of particle species. For the spectrum considered here, $N$ is sufficiently low that $\Lambda_{\rm sp}$ remains within $\sim 10 \% $ of $M_{\rm{Pl}}$ and still lies above the Landau poles.
\begin{table}[t]
\vspace{0.5cm}
\centering
\small
\renewcommand{\arraystretch}{1.15}
\begin{tabular}{c c c c c c c}
\toprule
state & mult. & mass & $Y$
& $q_\chi=(q^1,q^2,q^3)$
& $q_{\chi^c}=(q^1,q^2,q^3)$\\
\midrule
$1$  & $10$ & $\Phi_1$     & $+0$ & $(+0,+0,+2)$ & $(+1,+1,-1)$ \\
$2$  & $3$  & $\Phi_1$     & $+0$ & $(+0,-1,+0)$ & $(+1,+2,+1)$ \\
$3$  & $3$  & $\Phi_1$     & $+0$ & $(+1,+2,+3)$ & $(+0,-1,-2)$  \\
$4$  & $1$  & $\Phi_1$     & $+1$ & $(+1,-1,+2)$ & $(+0,+2,-1)$  \\
$5$  & $1$  & $\Phi_1$     & $+1$ & $(+1,-1,+3)$ & $(+0,+2,-2)$\\
$6$  & $1$  & $\Phi_1$     & $+1$ & $(+1,+2,+3)$ & $(+0,-1,-2)$ \\
$7$  & $1$  & $\Phi_1$     & $+0$ & $(+1,+3,+0)$ & $(+0,-2,+1)$  \\
$8$  & $6$  & $\Phi_1^{*}$ & $+0$ & $(+2,-2,-3)$ & $(-3,+1,+2)$ \\
$9$  & $5$  & $\Phi_1^{*}$ & $+0$ & $(+2,+0,-2)$ & $(-3,-1,+1)$  \\
$10$ & $4$  & $\Phi_1^{*}$ & $+0$ & $(-3,-1,+0)$ & $(+2,+0,-1)$  \\
$11$ & $3$  & $\Phi_1^{*}$ & $+0$ & $(+2,-3,+1)$ & $(-3,+2,-2)$  \\
$12$ & $2$  & $\Phi_1^{*}$ & $+0$ & $(+2,-1,+0)$ & $(-3,+0,-1)$ \\
$13$ & $2$  & $\Phi_2$     & $-1$ & $(-2,-3,-3)$ & $(-3,+3,+3)$ \\
$14$ & $1$  & $\Phi_2$     & $-1$ & $(-2,+3,+2)$ & $(-3,-3,-2)$ \\
$15$ & $2$  & $\Phi_2^{*}$ & $+1$ & $(+3,+0,+1)$ & $(+2,+0,-1)$ \\
$16$ & $1$  & $\Phi_2^{*}$ & $+1$ & $(+3,+3,+2)$ & $(+2,-3,-2)$ \\
$17$ & $1$  & $\Phi_3$     & $+1$ & $(+1,+0,+0)$ & $(-1,-1,+0)$ \\
$18$ & $1$  & $\Phi_3$     & $-1$ & $(+1,+2,-3)$ & $(-1,-3,+3)$  \\
$19$ & $1$  & $\Phi_3$     & $-1$ & $(+2,+1,+0)$ & $(-2,-2,+0)$ \\
$20$ & $1$  & $\Phi_3^{*}$ & $+0$ & $(+2,+3,-3)$ & $(-2,-2,+3)$ \\
$21$ & $1$  & $\Phi_3^{*}$ & $+0$ & $(+3,+2,+3)$ & $(-3,-1,-3)$  \\
$22$ & $1$  & $\Phi_3^{*}$ & $+0$ & $(+3,+3,+2)$ & $(-3,-2,-2)$  \\
\bottomrule
\end{tabular}
\caption{Example colorless canceller spectrum for the down-type star big axion: $22$ distinct types comprising $52$ SM-vectorlike pairs in total, with $\sum_I n_IY_I^2=12$. All spectator hidden charges satisfy $|q^a|\leq3$. The spectrum cancels the remaining gauge anomalies, contributes $\Delta b_Y^\chi=16$ to the one-loop hypercharge beta function, and contains five types with direct SM mixing portals of dimension six or below.}
\label{tb:star-cancellers}
\end{table}

These extra states must also preserve theory space non-locality -- as emphasized by Ref.~\cite{Bonnefoy:2022vop}, effective Coleman-Weinberg operators generated by fermion loops can in principle lower the quality of the big axion. However, within the spectator spectrum described by Table~\ref{tb:star-cancellers}, we find that the leading one-loop Coleman-Weinberg operator sensitive to the big axion is proportional to $\Phi_1^5\Phi_2\Phi_3^5\Phi_4^5$, corresponding to $D_{\rm quality}^{\rm eff}=16$. Assuming all higher-dimension operators are suppressed by the Planck scale, the big axion thus retains its quality after the introduction of the spectator states.

The inclusion of these new states raises their own cosmological questions if they are produced in the early universe. While not every mass spectrum is cosmologically safe, a viable decay texture can be obtained by placing the fast-decaying portal states---those which mix sufficiently with the SM---at the bottom of the heavy spectrum and allowing the remaining states to cascade into them. The spectrum contains five types with direct SM portals of dimension six or below, namely $1,\ 2,\ 4,\ 6$ and $17$ whose lowest portal dimensions are $6$, $5$, $6$, $6$, and $5$, respectively. Every remaining state can be connected to one of these portal states through gauge-invariant off-diagonal fermion mixings of dimension at most seven. Because the big axion is essentially massless, these mixings allow the heavier hidden fermions to cascade downward by axion emission, $\chi_a\rightarrow\chi_b+a$, before the portal states decay into SM particles.\footnote{One possible cascade structure is a mass hierarchy allowing for $3\to1$, $7\to2$, $20\to1$, $9\to20$, $11\to20$, $12\to9$, $10\to12$, $21\to2$, $8\to21$, and $22\to21$ in the hypercharge-neutral sector, together with $5\to4$, $18\to5$, $13\to6$, $19\to6$, $14\to19$, $15\to19$, and $16\to17$ in the charged sector.} A detailed treatment of the resulting spectator thermal history is left for future work.

\bibliography{big-bib}

\end{document}